\documentclass[usenatbib,fleqn]{mn2e}

\usepackage{graphicx,subfigure}
\usepackage{amsmath,amssymb}

\begin{document}

\title[The structure of H{\sc{i}} in galactic disks]
{The structure of H{\sc{i}} in galactic disks: Simulations vs observations}

\author[David M. Acreman et al]{David M. Acreman$^1$\thanks{ E-mail acreman@astro.ex.ac.uk}, Clare L. Dobbs$^{1,2,3}$, Christopher M. Brunt$^1$, Kevin A. Douglas$^4$ \\
$^1$ School of Physics, University of Exeter, Stocker Road, Exeter EX4
4QL.\\
$^2$ Max-Planck-Institut f\"ur extraterrestrische Physik, Giessenbachstra\ss{}e, D-85748 Garching, Germany \\
$^3$ Universitats-Sternwarte M\"unchen, Scheinerstra\ss{}e 1, D-81679
M\"unchen, Germany \\
$^4$ Arecibo Observatory/NAIC, HC 3 Box 53995, Arecibo, PR 00612  USA
}

\maketitle

\begin{abstract}

  We generate synthetic H{\sc{i}} Galactic plane surveys 
  from spiral galaxy simulations which include stellar feedback
  processes. Compared to a model without feedback we find an increased
  scale height of H{\sc{i}} emission (in better agreement with
  observations) and more realistic spatial structure (including
  supernova blown bubbles).  The synthetic data show H{\sc{i}}
  self-absorption with a morphology similar to that seen in
  observations. The density and temperature of the material
  responsible for H{\sc{i}} self-absorption is consistent with
  observationally determined values, and is found to be only weakly
  dependent on absorption strength and star formation efficiency.

\end{abstract}


\section{Introduction}

H{\sc{i}} emission, from the 21-cm line of atomic hydrogen, is
widely used as a tracer of large scale Galactic structure
\citep{Kalberla09} but also reveals smaller scale structure in the
interstellar medium. This smaller scale structure includes shells
formed by stellar feedback processes, which inject kinetic and thermal
energy, along with heavy elements. Such feedback is instrumental in
determining the evolution of a galaxy and its effects can be seen in
our own Galaxy \citep{Heiles79,Heiles84,Taylor03} and also in nearby
external galaxies \citep{Fukui09,Bagetakos11}. Hence feedback
processes are seen to play an important role in the formation of
smaller scale structure and its appearance in H{\sc{i}}
emission. Numerical simulations of the interstellar medium (ISM) in
disk galaxies are becoming increasingly sophisticated and can now
include the effects of stellar feedback processes \citep{Wada08,
  Shetty08, Dobbs11,Tasker11}. At the same time advances are being
made in the generation of synthetic observations from simulations
\citep[and references
therein]{Hennebelle07,douglas_2010,Parkin11}. These synthetic
observables allow a direct comparison between simulation results and
real observations, and as the models become increasingly sophisticated
we can expect the synthetic observations to compare more favourably
with real observations. As well as testing whether models can produce
realistic observables, such synthetic observations also allow features
to be related more directly to underlying physical processes than is
possible with real observations. Hence synthetic observations can be
used to test physical mechanisms proposed to explain features seen in
observations.

The transition of hydrogen gas from the atomic phase to the
    molecular phase is a key step to molecular cloud, and ultimately
    star formation. In order for this transition to occur the gas must
    become cooler and denser than typical ISM material. The passage of
    material through the shocks associated with the density waves in a
    grand design spiral galaxy results in substantial compression
    \citep{Roberts72}, which allows the gas to cool efficiently down
    to low temperatures \citep{Cowie81,Bergin04,DGCK08,Kim08}. 
    The high density, low temperature material in a spiral shock
        provides favourable conditions for the formation of molecular
        clouds (although molecular clouds can also form in the absence
        of spiral shocks).
    Whilst CO traces molecular
    clouds, cold H{\sc{i}} with a low molecular fraction is more
    difficult to detect. It can however be detected as H{\sc{i}}
    self-absorption (HISA). HISA is observed when colder, denser
    foreground material is located along the same line of sight, and
    at the same line of sight velocity, as warmer background material
    \citep{Li2003,Gibson05}. Given these requirements, using HISA is
    not a suitable method to obtain a complete map of cold H{\sc{i}}. However
    HISA can provide information in cases where other tracers of star
    forming clouds (such as the J=1--0 rotational transition of the CO
    molecule) are not effective \citep{Douglas07}, for example shortly
    after passing through a shock \citep{Bergin04}, or in photon
    dominated regions where there is less effective self-shielding of
    CO against the interstellar radiation field compared to $\rm{H}_2$
    \citep{Kaufman99,Allen04,Shetty11}. Thus although HISA is found to
    correlate with emission from molecular species
    \citep{Goldsmith05,Kavars05} the correspondence between HISA and
    tracers of star forming clouds is not straightforward. Synthetic
    observations provide a valuable way of examining tracers such as
    HISA in a more controlled environment.

In this paper we present synthetic H{\sc{i}} Galactic
plane surveys derived from simulations of galaxies which include
stellar feedback and compare the H{\sc{i}} emission
and HISA structure with similar features found in observations. We
begin in Section~\ref{section:method} by outlining the method used to
make the synthetic H{\sc{i}} surveys. In Section~\ref{section:effect}
we assess the effect of feedback on the observable properties of the
H{\sc{i}} gas by comparing our synthetic data with observational data
and with a previous synthetic survey which did not include
feedback. The distribution of HISA in our synthetic observations is
discussed in Section~\ref{subsection:HISA_distro} and the properties
of the material producing HISA are investigated in
Section~\ref{subsection:HISA_material}. We finish by presenting our
conclusions in Section~\ref{section:conclusions}.

\section{Method}
\label{section:method}

Our synthetic observations are generated from an SPH (smoothed
    particle hydrodynamics) simulation of a spiral galaxy. The variable
    spatial resolution of the SPH method permits a large range of
    density and spatial scales to be resolved. This enables a
    simulation with a domain covering a whole galaxy, which can still
    resolve individual molecular clouds. SPH is a Lagrangian method
    which means that the material associated with an SPH particle
    retains its identity as the simulation progresses. Consequently we
    can track material as it passes thorough a spiral arm and
    determine its fate at a later time. A synthetic Galactic plane
    survey was previously presented by \cite{douglas_2010} (hereafter
    Paper~1) using an SPH model without feedback processes. The new
    results presented here include stellar feedback processes in the
    SPH model, which was proposed as an important mechanism to resolve
    differences between Paper 1 results and observations.

The SPH simulations used for this work are models C and D from
\cite{Dobbs11} which include stellar feedback and self-gravity of
    the gas, which were not included in the simulation of
\cite{dobbs08} used in Paper~1. In all the simulations the gas is
    assumed to orbit in a fixed potential, composed of halo, disc and
    four armed spiral components. The resulting rotation curve is
    comparable to that of the Milky Way, which allows structures in
    the simulations to be compared to observed structures in our own
    Galaxy. For example the simulations have spiral arms which are
    similar to those in our Galaxy which allows us to generate
    synthetic images analogous to observations of the nearby Perseus
    arm in our Galaxy. The total surface density (including Helium) is
    constant with values of $8~\rm{M}_{\sun}pc^{-2}$ for models
    including feedback and $10~\rm{M}_{\sun}pc^{-2}$ for the model
    without feedback used in Paper~1. This is comparable to the
    surface density in the solar neighbourhood \citep{Wolfire03}.

Stellar feedback is inserted as kinetic and thermal
energy, where gas is assumed to have formed stars, although sink
    or star particles are not introduced.
For feedback to occur the density of a gas particle in a
converging flow must exceed 1000 $\rm{cm}^{-3}$, whilst the
surrounding region of gas (within a radius of $\sim20$~pc) must be
gravitationally bound and in an energetically favourable state
\citep{Dobbs11b}. The number of supernovae is estimated from the mass
of gas in the computed region, adopting a certain star formation
efficiency and a Salpeter initial mass function (IMF). 
For each feedback event, the total amount of energy deposited is
    given by equation 1 from \cite{Dobbs11b}, which computes the
    number of massive stars expected to form, and assumes each
    supernova injects $10^{51}$ erg of energy. 
    The number of stars formed (and therefore the amount of energy
    deposited) depends on a star formation efficiency parameter,
    $\epsilon$, which represents the fraction of the molecular mass in
    each bound region assumed to be turned into stars.
    The star formation efficiency is an absolute value and does not 
    depend on the free-fall time. The energy is distributed
    according to a snowplough solution, which describes the
    pressure-driven phase of a supernova after the blast wave
    \citep{Woltjer72,Chevalier74,McKee77} and is deposited as 2/3
    kinetic energy, and 1/3 thermal energy.
The feedback is assumed to be instantaneous,
and although we consider supernova explosions, the energy could
account for numerous feedback processes, such as stellar winds,
radiation and supernovae. 

Model data at a simulated time of 250~Myr are used, at which time the
simulated galaxy has reached a state of quasi-equilibrium. In
    Model~C approximately one third of the ISM is in each of the cold
    ($<150\rm{K}$), unstable (150--5000K) and warm ($>5000\rm{K}$)
    phases. With a higher star formation efficiency there is more
    material in the warm phase and less material in the cold
    phase. The thermal evolution of the gas is modelled according to
    the thermodynamics of \cite{DGCK08}. The gas exhibits a maximum
    temperature of $2 \times 10^{6}$K. A lower limit of 20K is imposed
    prior to the temperature update in the model. This limit prevents
    material from becoming too cold for the simulation to treat
    accurately but allows some cooling below 20K to occur. The
    simulation follows the evolution of molecular gas, although only a
    small fraction ($\lesssim$ 10 percent) is molecular. Energy is
    injected as soon as the criteria for feedback are met, which
    results in dense gas being rapidly disrupted. Including a delay
    before injecting energy from feedback would enable longer for
    molecular gas to form.

The SPH data provide three dimensional distributions of H{\sc{i}}
density, temperature and velocity which are used to set up an adaptive
mesh refinement (AMR) grid for the {\sc{torus}} radiative transfer
code \citep{Harries00}. {\sc{torus}} carries out non-parallel ray
traces (one per image pixel and velocity channel) to solve the
radiative transfer equation and generate three-dimensional spectral
data cubes of 21~cm H{\sc{i}} emission in Galactic
longitude-latitude-velocity co-ordinates.

The AMR grid is generated by splitting the grid cells so that there is
never more than one SPH particle per cell. Mapping temperature,
density and velocity onto the AMR grid is carried out as described in
Paper~1 and \cite{acreman_2010b} using the method of
\cite{rundle_10}. The observer is placed within the model galaxy, at a
location analogous to that of the Sun within our own Galaxy, and the
synthetic survey is generated as if it were the Galactic second
quadrant ($90^{\circ}<l<180^{\circ}$). The ray tracing uses the
density sub-sampling method of \cite{rundle_10} to linearly
interpolate density values within a cell on the AMR grid. Density
sub-sampling was not required in Paper~1 as the spatial resolution of
the SPH simulation and AMR grid were finer (the computational demands
of adding feedback processes require that a lower spatial resolution
is used). The data cubes have a velocity resolution of 0.5~km/s and
an angular resolution of 1~arcmin in longitude and latitude.

Each cell on the AMR grid is assigned a thermal line width, which does
not include any additional turbulent component to account for
unresolved structure. This avoids adding an ad-hoc parameter but will
tend to under-estimate the line width if unresolved structure makes a
significant contribution to the velocity dispersion
\citep{Hennebelle07}. The velocity dispersion of the gas and clouds in the 
simulations are discussed in \cite{Dobbs11}.

\section{Effect of feedback on H{\sc{i}} structure}
\label{section:effect}

\subsection{Latitude-longitude structure}
\label{subsection:lb_structure}

\begin{figure*}
  \begin{center}
  \subfigure[NoFeedback]{
  \includegraphics[scale=1.1]{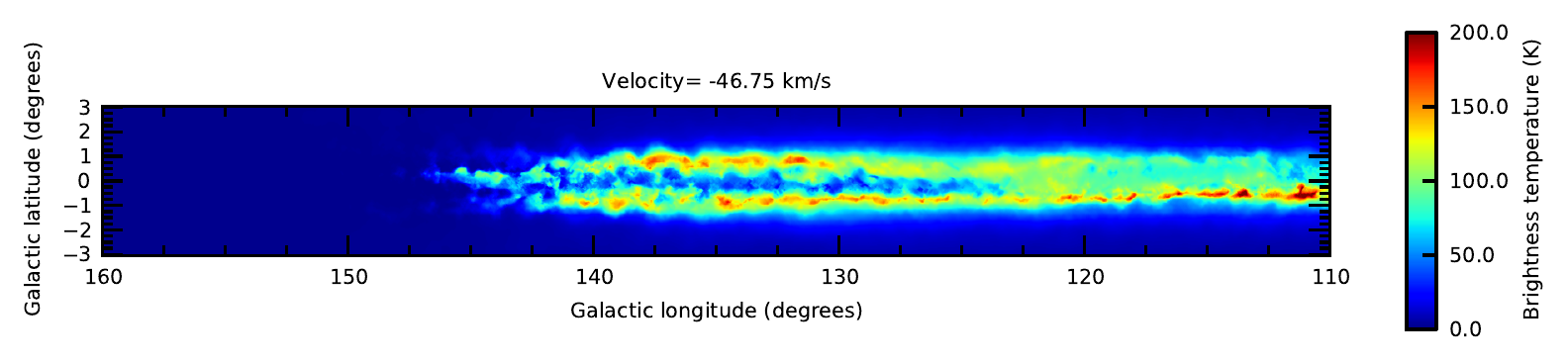}\label{fig:GHI_p2}}
\subfigure[Feedback5]{ 
  \includegraphics[scale=1.1]{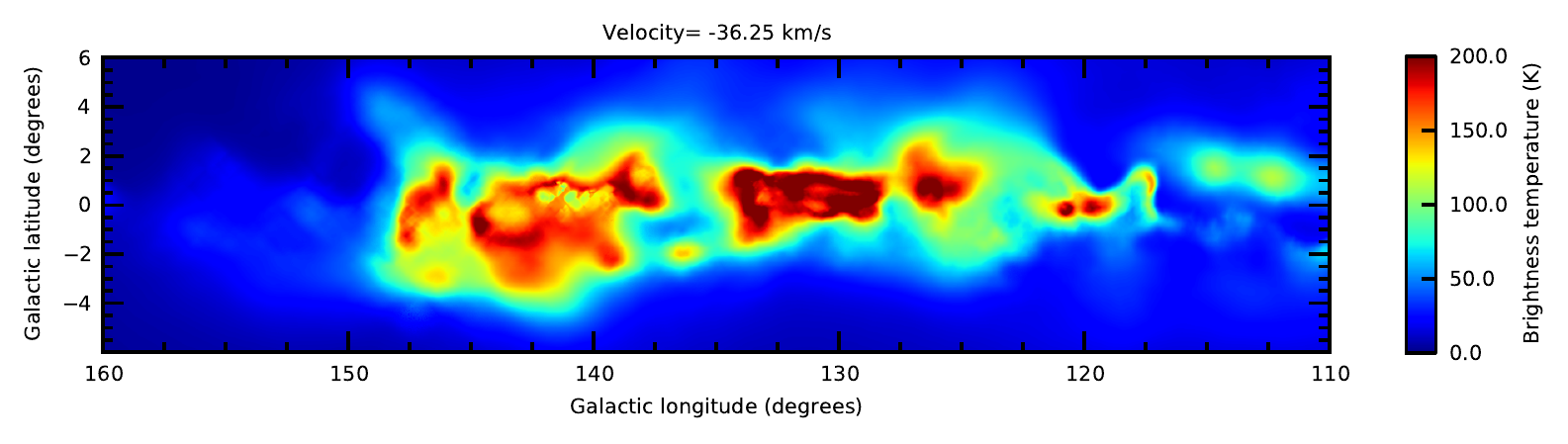}\label{fig:GHI_fb}}
\subfigure[Feedback10]{
  \includegraphics[scale=1.1]{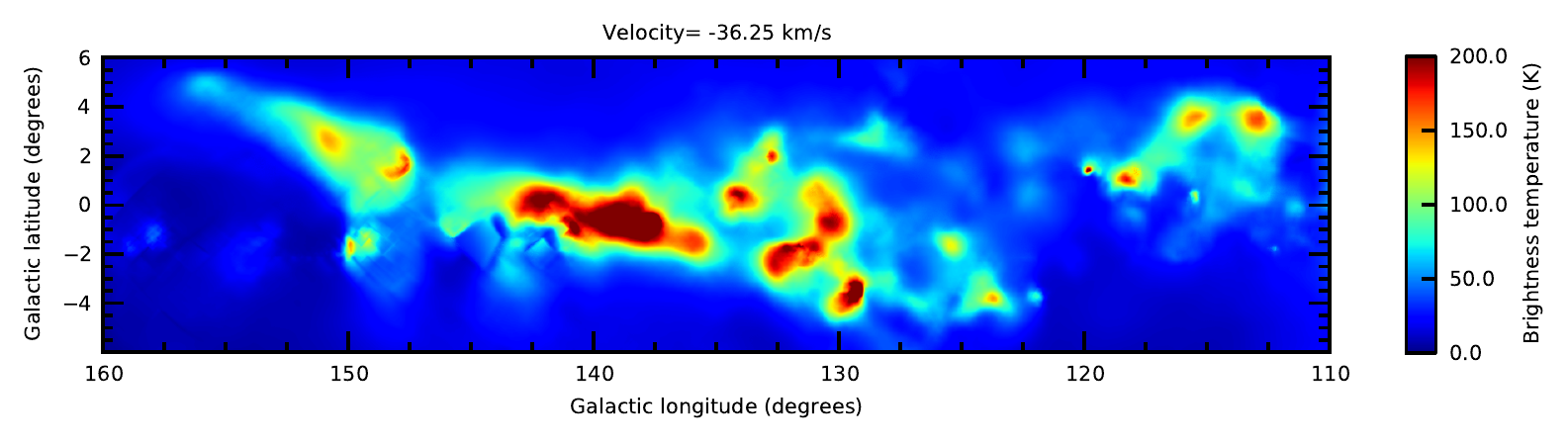}\label{fig:SQC}}
\subfigure[CGPS]{
  \includegraphics[scale=1.1]{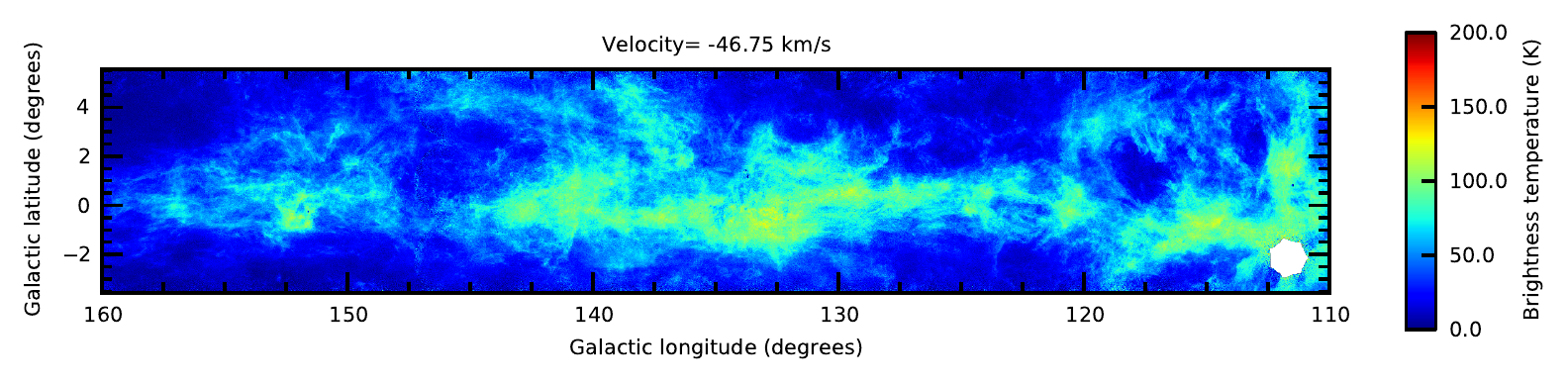}\label{fig:GHI_cgps}}
\caption{Brightness temperature of H{\sc{i}} emission, in Galactic
  latitude and longitude co-ordinates, from
  the Perseus arm in the CGPS and Perseus arm analogues in the
  synthetic data. The longitude coverage is the same in each case but
  the latitude coverage varies according to the extent of H{\sc{i}}
  emission (synthetic data) or survey coverage
  (CGPS). Fig.~\ref{fig:GHI_p2} is from a model galaxy without
  feedback, Fig.~\ref{fig:GHI_fb} is from a model galaxy with feedback
and 5~per cent star formation efficiency, Fig.~\ref{fig:SQC} is from a model galaxy with feedback
and 10~per cent star formation efficiency, and Fig.~\ref{fig:GHI_cgps}
is from the Canadian Galactic Plane Survey.}
  \label{fig:GHI}
\end{center}
\end{figure*}

Figure~\ref{fig:GHI} shows latitude-longitude plots of H{\sc{i}}
emission for simulations with and without feedback, and also for the
Canadian Galactic Plane Survey (CGPS) observations
\citep{Taylor03}. The CGPS data are from a velocity channel which
contains emission from Perseus arm material. In the synthetic
observations the observer has been positioned so that at similar
velocities we see a spiral arm analogous to the Perseus
arm. Fig.~\ref{fig:GHI_p2} shows synthetic observations from a
simulation without feedback from Paper~1 (hereafter referred to as
NoFeedback). Figure~\ref{fig:GHI_fb} is from Run~C of \cite{Dobbs11}
and has feedback included with 5\% star formation efficiency
(hereafter referred to as Feedback5), and Fig.~\ref{fig:SQC} is from
Run~D of \cite{Dobbs11} and has feedback included with 10\% star
formation efficiency (hereafter referred to as Feedback10). The
    star formation efficiency referred to here is the $\epsilon$
    parameter described in Section~\ref{section:method}.

The CGPS data show a broad region of emission which, in places,
extends beyond $\pm2$~degrees outside the mid-plane. In contrast the
NoFeedback run has emission which is much more confined to the
mid-plane, with bright ridges of emission at $\pm1$~degree which are
not seen in the CGPS data. Without feedback the gas in the model
    galaxy is overly confined to the mid-plane, resulting in large
    mid-plane optical depths. In the NoFeedback model the column
    density is highest within $\pm1$~degree of the mid-plane but the
    accumulation of cold, dense material results in excessively high
    absorption and correspondingly low levels of H{\sc{i}} emission
    (in the optically thick limit the brightness temperature will
    saturate at the spin temperature).  When feedback is included
    (Fig.~\ref{fig:GHI_fb} and \ref{fig:SQC}) the gas in the model
    galaxy is much less confined to the mid-plane and the bright
    ridges of emission, seen in Fig.~\ref{fig:GHI_p2} at about
    1~degree above and below the mid-plane, are not present. The
    H{\sc{i}} emission now extends further out of the mid-plane, in
    better agreement with the CGPS observations.

Feedback results in significant holes in H{\sc{i}} emission, with
Feedback10 in particular having much less contiguous emission than the
CGPS observations. The CGPS observations have small scale filamentary
structure, which is not seen in the synthetic observations, however,
we do not expect to reproduce this structure at the current SPH
resolution. The resolution of the SPH simulation is governed by
    the smoothing lengths of the particles\footnote{ The smoothing
      length $h$ is given by
\begin{equation}
h=\eta \left( \frac{m}{\rho} \right)^{1/3} \nonumber
\end{equation}
where $m$ is the particle mass, $\rho$ is the particle density, and 
$\eta=1.2$ \citep{Price07}.}.
The density threshold for feedback to occur is at a
number density of $1000\rm{cm}^{-3}$ which corresponds to a smoothing
length of 5.6~pc (with a particle mass of $2500\rm{M}_\odot$). At a
distance of 2.5~kpc (typical of material seen in Fig.\ref{fig:GHI})
this corresponds to an angular size of 0.13 degrees (or
7.8~arcmin). Structure on this scale is seen in the synthetic data
(e.g. at $l=120$ in Fig~\ref{fig:SQC}) but is spherical, rather than
filamentary.

%

\subsection{Vertical distribution of H{\sc{i}} emission}

In order to allow a quantitative comparison of the vertical
distribution of H{\sc{i}} emission, longitudinally averaged profiles
of brightness temperature against latitude were extracted for the
longitude range 126--144~degrees, in the same velocity channels shown
in Fig.~\ref{fig:GHI}. Emission seen in this longitude and velocity
range is from spiral arm material at a distance of approximately 2.5~kpc
from the observer. At this distance material at a latitude of 1~degree
is 44~pc out of the mid-plane. 

\begin{figure*}
\centering
\subfigure[Latitude axis not inverted]{
  \includegraphics[scale=0.31]{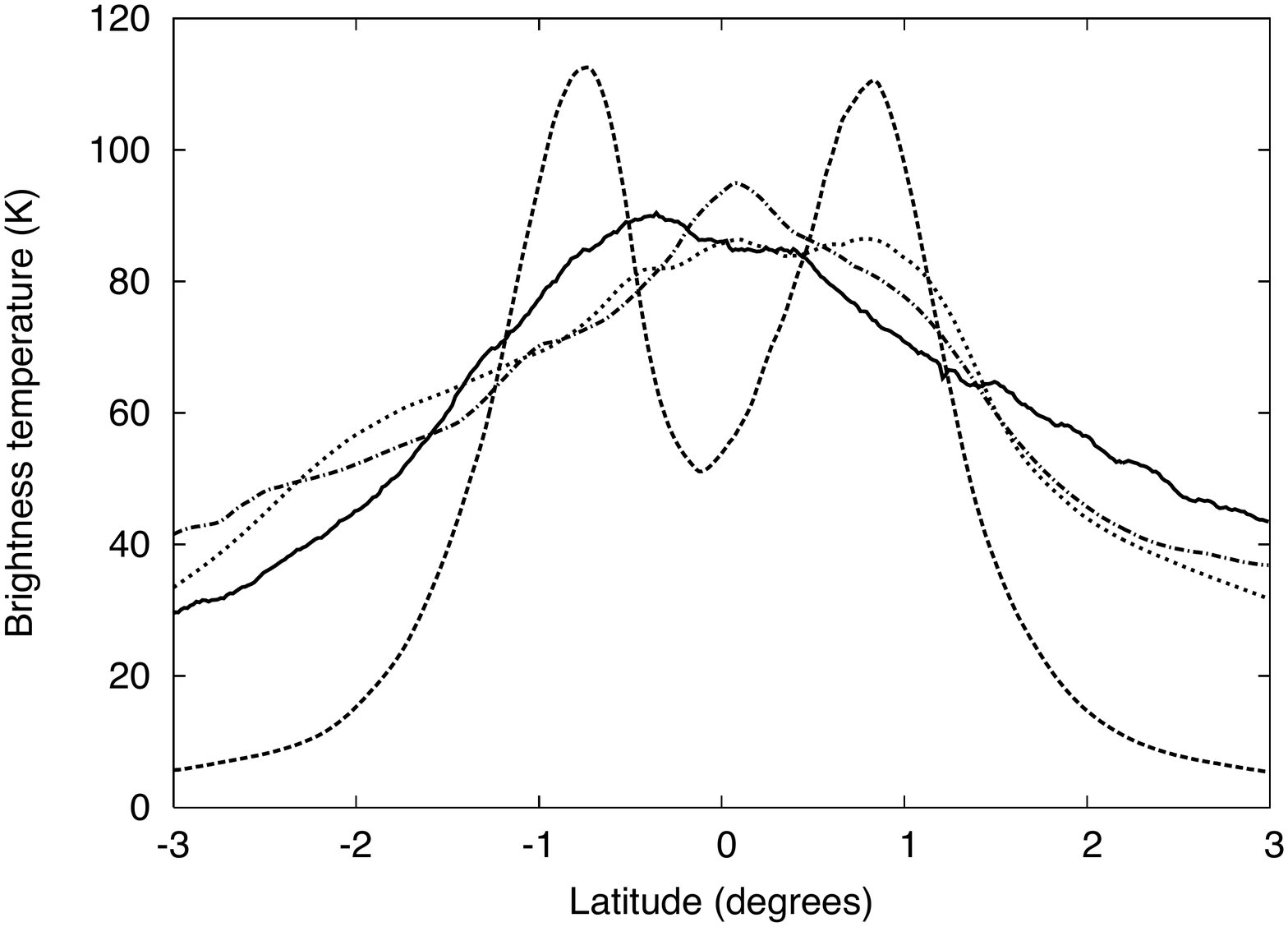}\label{fig:scale_height_profs_norev}}
\subfigure[Latitude axis inverted]{
  \includegraphics[scale=0.31]{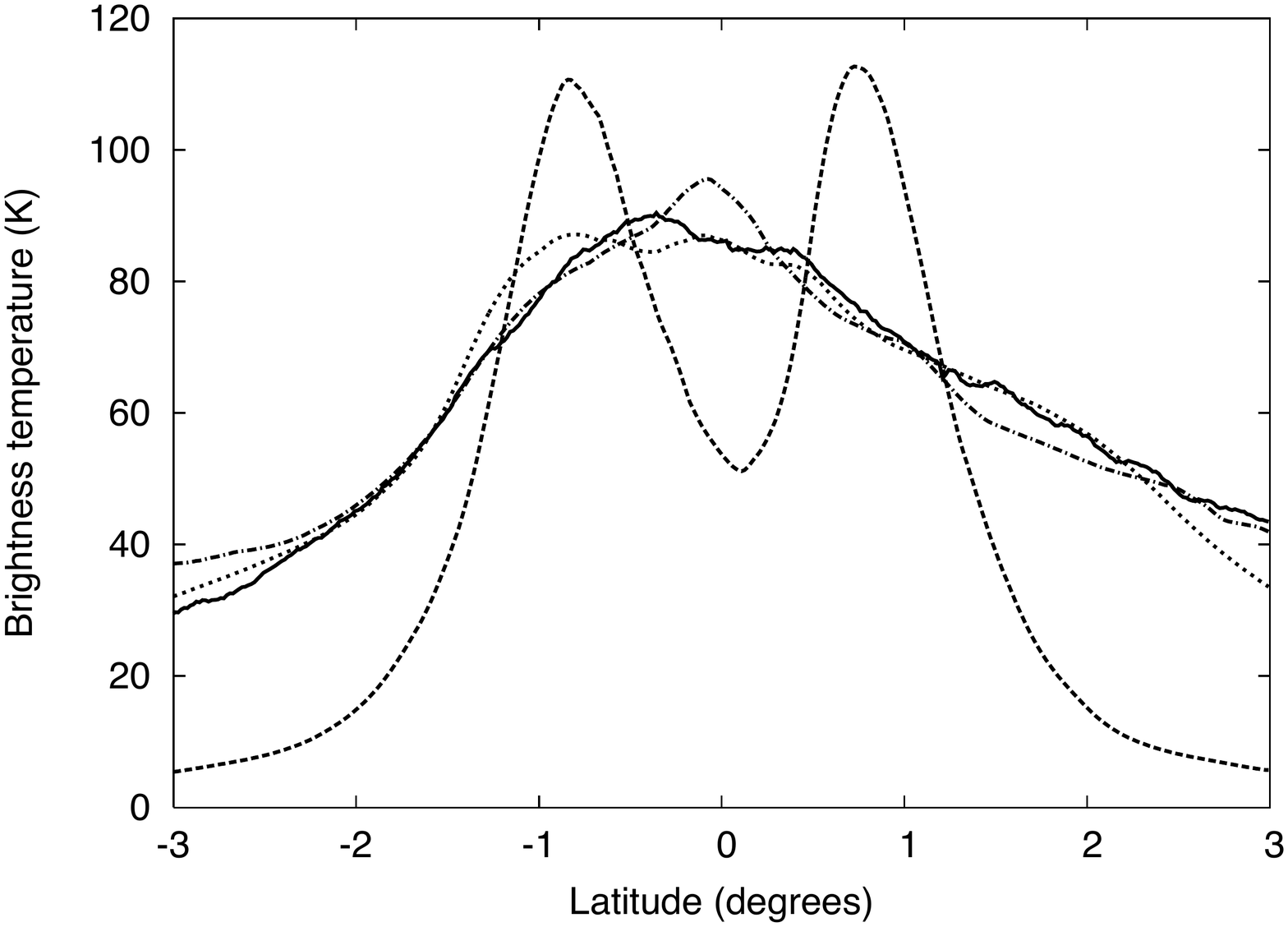}\label{fig:scale_height_profs_rev}}
\caption{Profiles of H{\sc{i}} brightness temperature with
  latitude. The solid line is from the CGPS, the dashed line is from
  the NoFeedback model, the dotted line is from Feedback5 and the
  dot-dashed line is from Feedback10. The normalisation of the
  synthetic profiles has been scaled to fit the CGPS data, so that the
  shape of the profiles can be readily compared. In
  Fig.~\ref{fig:scale_height_profs_rev} the latitude axis is inverted
  so that the synthetic profiles better match skewness of the CGPS
  profiles.}
\label{fig:scaled_profiles}
\end{figure*}

The spiral arms in the simulations have a narrow velocity width,
    compared to the observed Perseus arm, and this results in emission
    from the simulated spiral arms being distributed over fewer
    velocity channels but with increased brightness temperatures (the
    structure of the arms in longitude-velocity space is discussed
    further in Section~\ref{subsection:l-v}). Consequently the
    synthetic surveys have higher brightness temperatures than the
    CGPS data and the shape of the raw profiles cannot easily be
    compared.  To facilitate a quantitative comparison the synthetic
profiles were scaled by a constant scaling factor chosen to minimise
the RMS difference from the CGPS data.  The expression used to find
the best value of the scaling factor $f$ was
\begin{equation}
\rm{RMS} = \sqrt{ \frac{1}{N}\sum_{i=1}^N \left( C_i - f S_j \right)^2 }
\label{eqn:rms}
\end{equation}
where $N$ is the number of points in the CGPS profile, $C_i$ is the
brightness temperature in bin $i$ of the CGPS profile, and $S_j$ is
the brightness temperature in bin $j$ of the synthetic profile, where
bin $j$ and bin $i$ correspond in latitude. Every point in the
brightness temperature profile was then scaled by the value of $f$
which minimised the expression in Eqn.~\ref{eqn:rms}. As the profiles
are asymmetric the fit was repeated with the latitude axis
inverted. The scaling factors and resultant RMS differences from the
CGPS profile are shown in Table~\ref{tab:scale_factors}. 
\begin{table}
  \centering
  \caption{Scaling factors applied to latitude profiles from synthetic
    data and the corresponding RMS differences from the CGPS profile.}
  \begin{tabular}{llll}
\hline
  Simulation & Scale factor & RMS  & Reversed axis \\
                    &    & difference (K) & axis? \\
\hline
  NoFeedback & 0.82 & 28.6 & No \\ 
  Feedback5   & 0.535 & 8.86 & No \\ 
  Feedback10   & 0.762 & 8.20 & No \\ 
\\
  NoFeedback & 0.821 & 28.5 & Yes \\ 
  Feedback5   & 0.539 & 3.39 & Yes \\ 
  Feedback10  & 0.767 & 3.51 & Yes \\ 
\hline
  \end{tabular}
  \label{tab:scale_factors}
\end{table}
The scaled profiles and the unmodified
CGPS profile are plotted in Fig.~\ref{fig:scale_height_profs_norev} and
in Fig.~\ref{fig:scale_height_profs_rev} (with a reversed latitude
axis in the second plot).

The fit for both the feedback runs is significantly better when the
latitude axis is reversed, as the skewness of the profile better
matches the skewness of the CGPS data.  Both Feedback5 and Feedback10 fit the
CGPS data well, provided the normalisation of the profile is scaled,
and it is not possible to infer from these results whether 5\% or 10\%
efficiency is a better fit to the data. Conversely the NoFeedback model does
not fit the CGPS data well as the profile shows a large dip around the
mid-plane giving a qualitatively different profile to that
observed. 

Although the star formation efficiency is different in Feedback5 and
Feedback10, the rate of star formation is comparable at 250~Myr, as
the process is largely self-regulating (as shown in fig.~11 of
\citealt{Dobbs11b}). Consequently the energy input from stellar
feedback is similar for Feedback5 and Feedback10 (as the supernova
rate is proportional to the star formation rate) and we expect the
impact on the vertical distribution of H{\sc{i}} to be
similar. Conversely a synthetic survey from a simulation with twice the
surface density (run M10 from \citealt{Dobbs11b}, not shown here) has
a larger scale height than Feedback5 and Feedback10. This is
consistent with the higher star formation rate (approximately four
times higher) and an increased level of stellar feedback.

The amount of energy injected into the ISM in the simulations is
    directly proportional to the star formation rate. The star
    formation rate (SFR) in the Feedback5 and Feedback10 models is
    $\sim0.7 {\rm{M}}_\odot/{\rm{year}}$, which is consistent with the
    lower limit of the observationally determined Galactic star
    formation rate of \cite{Robitaille10} but approximately a factor 2
    less than the SFR of \cite{Murray10}.  We note that SFRs
    determined from both observations and models are sensitive to
    assumptions about the slope and limits of the IMF
    \citep{Calzetti09}. Additionally we can only insert feedback on a
    resolvable scale in the simulations, and cannot take into account
    all the sub-resolution processes which may dissipate or radiate
    energy. Hence we cannot definitively say how well energy injection
    in the models is similar to that in the Milky Way. However the
    effect of feedback has clearly been to make the models more
    realistic in this regard.

\subsection{Longitude-velocity structure}
\label{subsection:l-v}

\begin{figure*}
  \centering
  \subfigure[NoFeedback]{
  \includegraphics[scale=1.0]{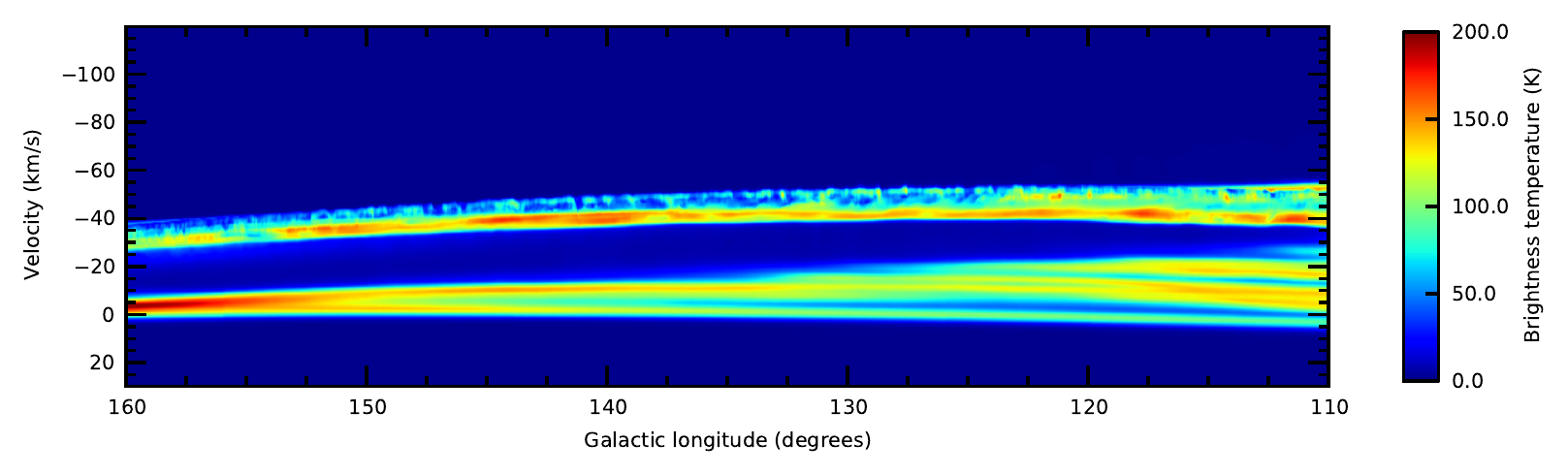}
\label{fig:l-v_NoFeedback}}
\subfigure[Feedback5]{
  \includegraphics[scale=1.0]{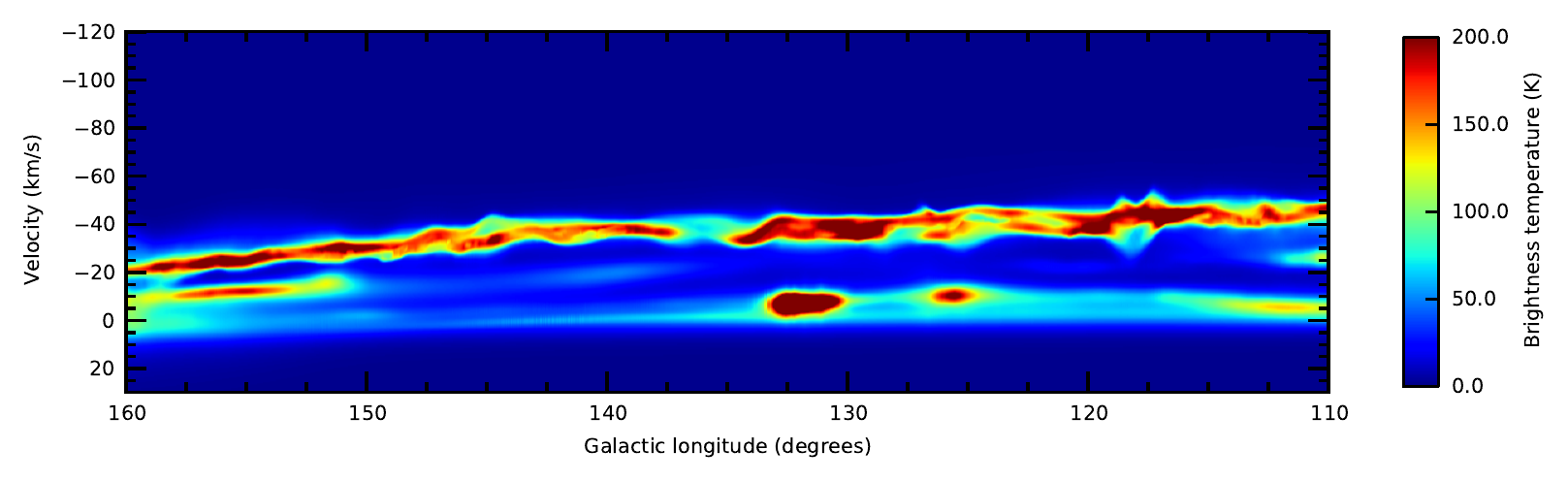}
\label{fig:l-v_Feedback5}}
\subfigure[Feedback10]{
  \includegraphics[scale=1.0]{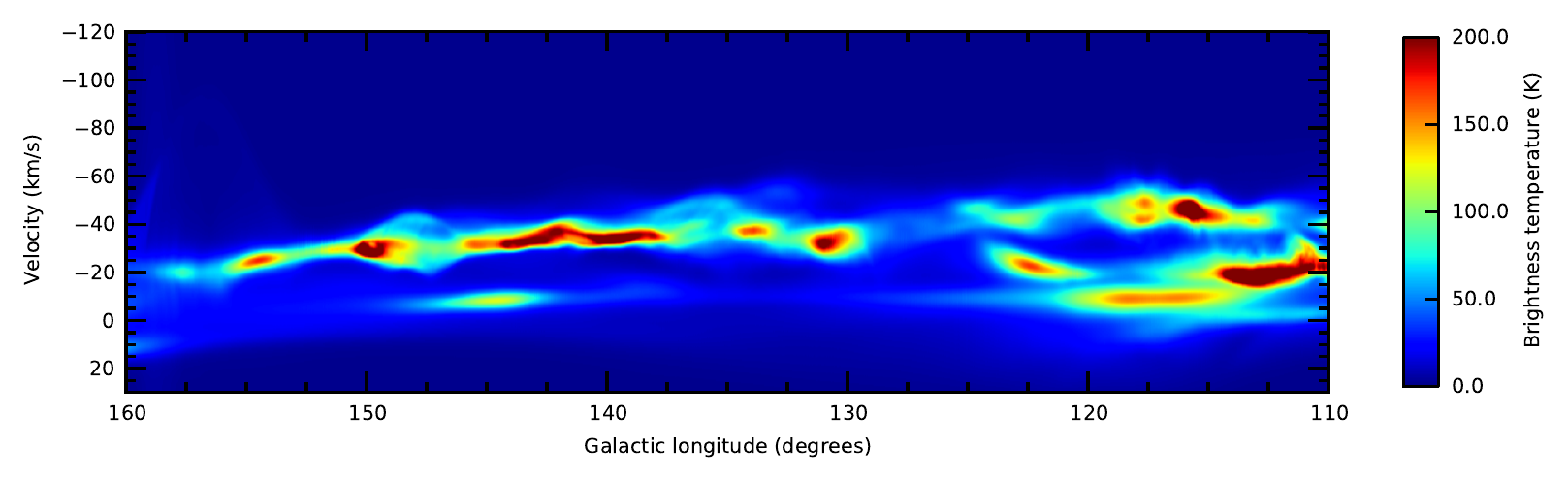}
\label{fig:l-v_Feedback10}}
\subfigure[CGPS]{
  \includegraphics[scale=1.0]{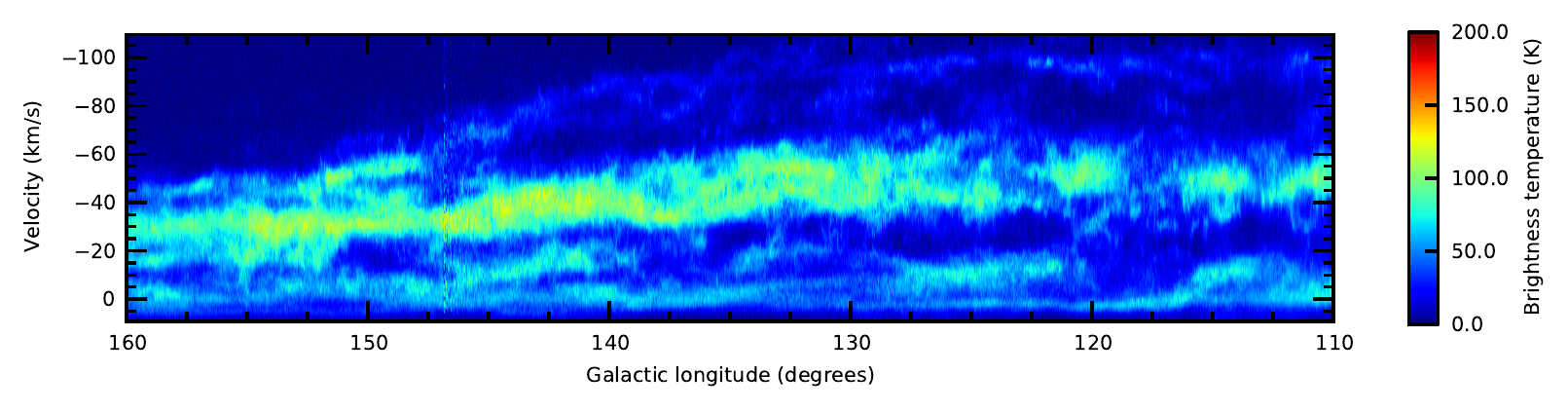}
\label{fig:l-v_CGPS}}
  \caption{Brightness temperature of H{\sc{i}} emission, in Galactic
  longitude-velocity co-ordinates in the
    Galactic mid-plane.  Fig.~\ref{fig:l-v_NoFeedback} is from a model galaxy without
  feedback, Fig.~\ref{fig:l-v_Feedback5} is from a model galaxy with feedback
and 5~per cent star formation efficiency, Fig.~\ref{fig:l-v_Feedback10} is from a model galaxy with feedback
and 10~per cent star formation efficiency, and Fig.~\ref{fig:l-v_CGPS}
is from the Canadian Galactic Plane Survey.}
  \label{fig:l-v}
\end{figure*}

Longitude-velocity plots in the Galactic mid-plane are shown in
Fig.~\ref{fig:l-v}. 
These plots show local material around v=0 and Perseus arm
    material at more negative velocities (between $-20$ and $-60$~km/s).
The inclusion of feedback results in much more structure in the spiral
arms in longitude-velocity space. In Feedback5 the Perseus arm
material is contiguous but structured, whereas in Feedback10 the
Perseus arm material appears less contiguous. The local material
    is highly disrupted by feedback but as this material is close
    to the observer the angular size of the SPH smoothing length is
    large and the material is poorly resolved.

The CGPS data show material at more negative velocities than the
simulations (beyond $-60$~km/s) but we would not expect to see
material in this velocity range in the synthetic data because the
model galaxies do not have a sufficiently large radial extent to
represent this material. 
For gas in axisymmetric circular rotation the line of sight velocity 
(from equation~1 of \cite{Kalberla09}) is 
\begin{equation}
v \left(R,z\right) = \left[ \frac{R_\odot}{R} \Theta \left(R,z\right)
  - \Theta_{\odot} \right] \sin l \cos b
\end{equation}
where $v \left(R,z\right)$ is the line of sight velocity at point
$\left(R,z\right)$, $\Theta\left(R,z\right)$ is the tangential
velocity at $\left(R,z\right)$, $\Theta_{\odot}$ is the observer's
tangential velocity, $R_\odot$ is the distance of the observer from
the Galactic centre, and $l$ and $b$ are Galactic latitude and
longitude. In our synthetic surveys the observer is located at
$R_\odot=7.1~\rm{kpc}$ within a model galaxy with an outer extent of
10~kpc, and at both these locations the tangential velocity is close to
220 km/s. Hence for material in the Galactic plane ($b=0$) the line of sight
velocity at the outer edge of the model galaxy is
\begin{equation}
  v \left(R,z\right) = -63.8 \sin l ~\rm{km/s}
\end{equation}
At $l=110$ the maximum extent in velocity space is expected to be
$-60~{\rm{km/s}}$ decreasing to $-22~{\rm{km/s}}$ at $l=160$.  This
is in agreement with the most extreme negative velocities seen in
the synthetic data in Fig.~\ref{fig:l-v}.

\subsection{Expanding shells}

\begin{figure*}
  \centering
\subfigure[Shell in longitude-latitude space]{
  \includegraphics[scale=0.475]{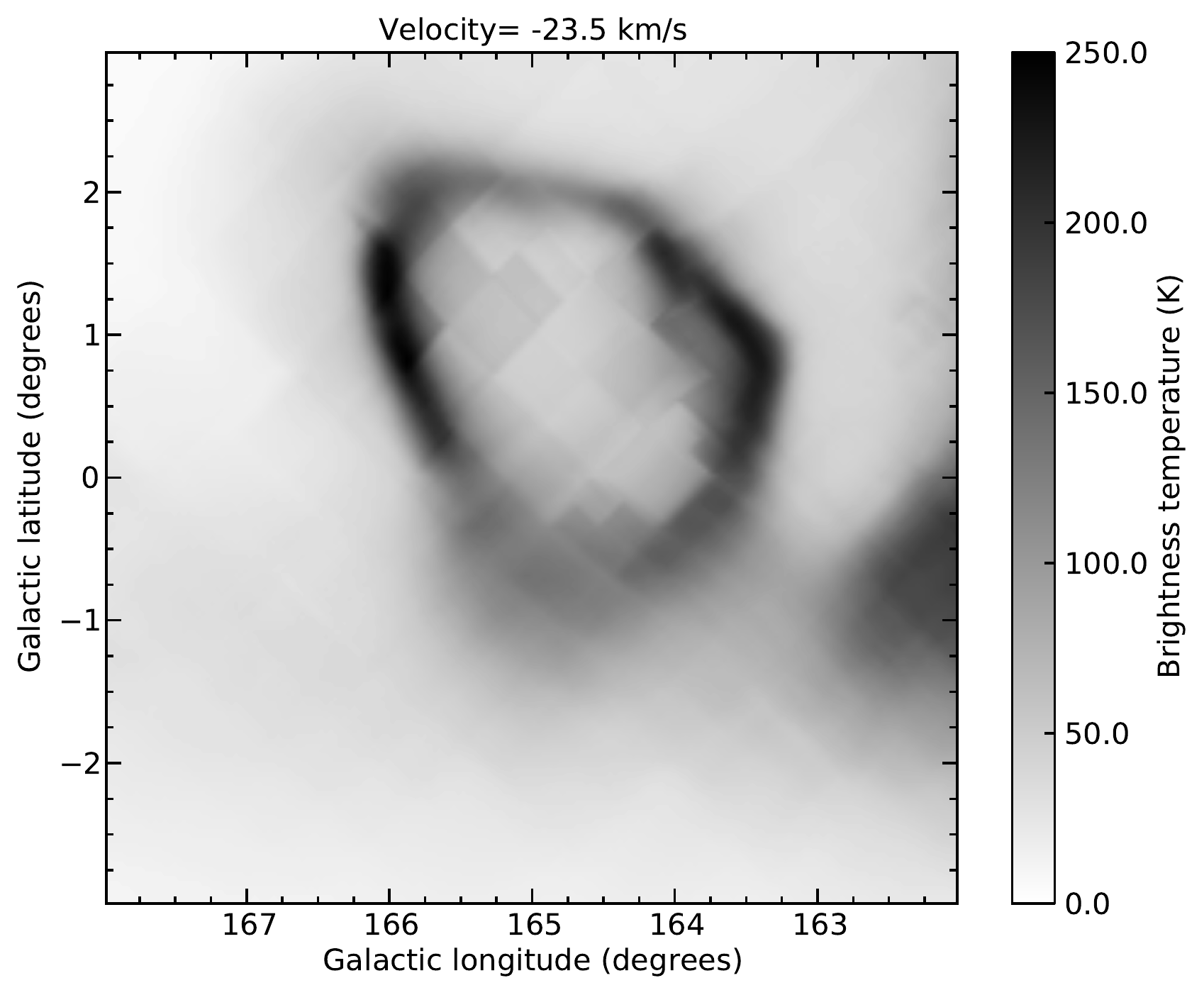}
  \label{fig:bubble_lb}}
\subfigure[Shell in velocity-radius space]{
  \includegraphics[scale=0.475]{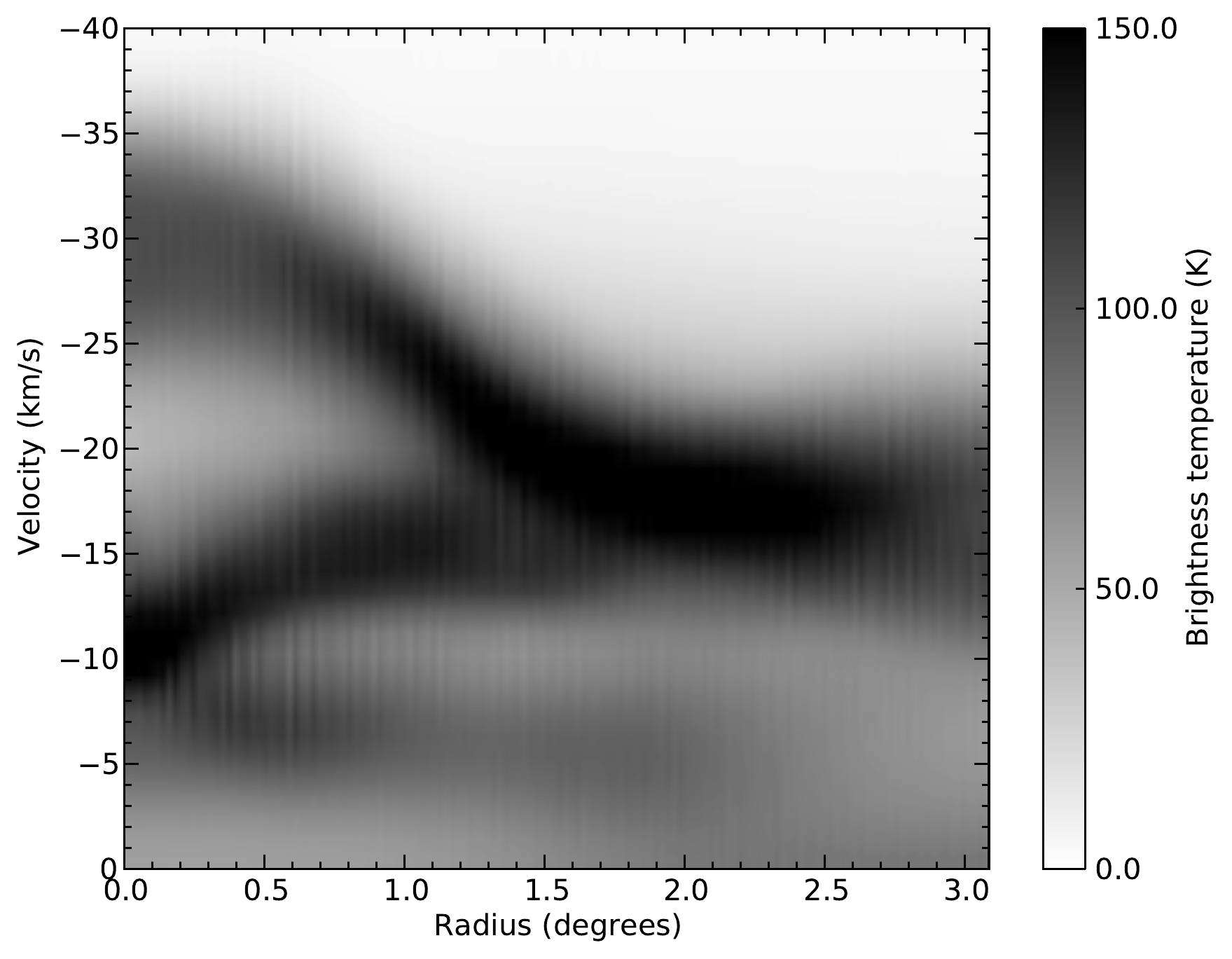}
  \label{fig:bubble_vr}}
  \caption{An expanding shell of H{\sc{i}} emission from the Feedback5
    model. The velocity-radius plot (right) shows the arch shape which is
    expected from an expanding shell. The angular offset is the
    radius of an annulus about the centre of the cell,
    where the centre is determined by eye.}
  \label{fig:bubble}
\end{figure*}

Several shells of material are seen in the synthetic observations (see
Fig.~\ref{fig:bubble} for an example from the Feedback5 model),
similar to shells seen in H{\sc{i}} observations \citep{Heiles79,
      Heiles84,Hu81,McClure-Griffiths02,Ehlerova05}.  Such structures
    are expected as a consequence of energy feedback from massive
    stars and the feature in Fig.~\ref{fig:bubble} is associated with
    a feedback event which occurred 1.2~Myr earlier\footnote{This time
      does not include the age of the supernova bubble when the
      energy is inserted. This timescale (denoted $t$ in Appendix~1,
      \citealt{Dobbs11b}) is of order $10^5$ years.}. The feedback event
    comprised 20 supernovae, an atypically energetic event, which
    injected a total of $2\times10^{52}$~erg of energy into the ISM.

In longitude-latitude space (Fig.~\ref{fig:bubble_lb}) an
    expanding shell appears as a variable radius ring with a
maximum radius at the central velocity of the material. The radius of
the shell decreases in velocity channels away from the central value
and terminates with two ``caps'' of emission from the front and back
of the bubble. By calculating the average surface brightness in
concentric annuli about the centre of the shell, the structure can be
plotted in velocity-radius space. A plot of this type, made using the
{\sc{kshell}} tool from the {\sc{karma}} software
package\footnote{http://www.science-software.net/karma/}, is shown in
Fig.~\ref{fig:bubble_vr}. In velocity-radius space the structure is an
arch shape which confirms that this structure is indeed expanding.

Figure~\ref{fig:bubble_vr} shows that the shell has an expansion
    velocity of approximately 10 km/s and a maximum angular radius of
    approximately 2~degrees, which corresponds to 80~pc at 2.3~kpc
    (the distance from the observer's position). The shell radius and
    expansion velocity are smaller than typical values from the
    samples of \cite{Heiles79} and \cite{Heiles84}, however
    \cite{Heiles84} notes a bias towards selecting larger shells and
    our shell is more typical of the shells found by \cite{Hu81}. The
    size of our shell is also consistent with the smaller shells in
    the sample of \cite{McClure-Griffiths02}, observed in the Southern
    Galactic Plane Survey.

\section{H{\sc{i}} self-absorption}
\label{section:hisa}

When generating synthetic observations the calculation of
    H{\sc{i}} intensity can be split into separate emitting and
    absorbing components. This allows us to produce a data cube
    containing only the absorption component, in order to identify
    where HISA is present. Furthermore, each cell of the AMR grid is
    identified as either a net source of emission or a net source of
    absorption by calculating the change in intensity due to the grid
    cell, normalised by the column density of the cell. The SPH
    particles are then associated with the change in emission from the
    AMR cell in which they reside, thus we are able to determine which
    SPH particles are responsible for HISA. Compared to observers we
    are in the privileged position of being able to unambiguously
    identify absorbing components in the data cube and furthermore we
    can identify absorption with specific SPH particles. In
    Section~\ref{subsection:HISA_distro} we use the
    absorption-only data cubes to examine the distribution of HISA on
    the plane of the sky, then in
    Section~\ref{subsection:HISA_material} we use the
    identification of absorbing SPH particles to investigate the
    properties of the material which causes HISA.

\subsection{HISA distribution}
\label{subsection:HISA_distro}

\begin{figure*}
  \centering
  \subfigure[NoFeedback]{
  \includegraphics[scale=1.1]{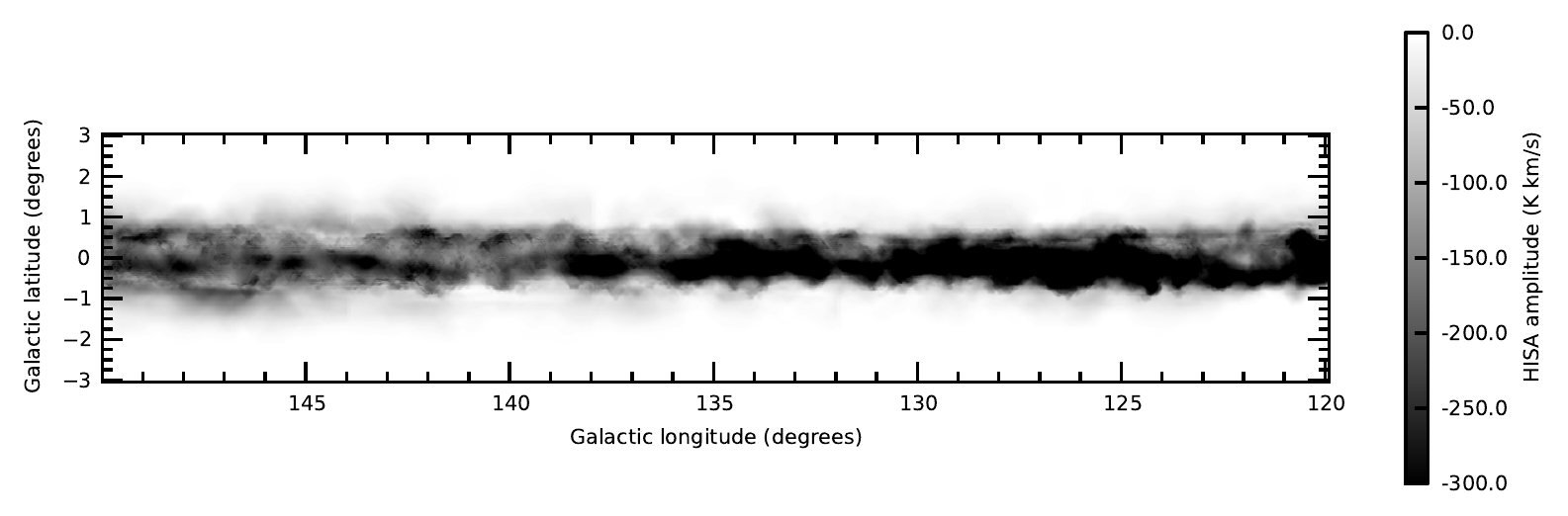}\label{fig:HISA_NoFeedback}}  
  \subfigure[Feedback5]{
  \includegraphics[scale=1.1]{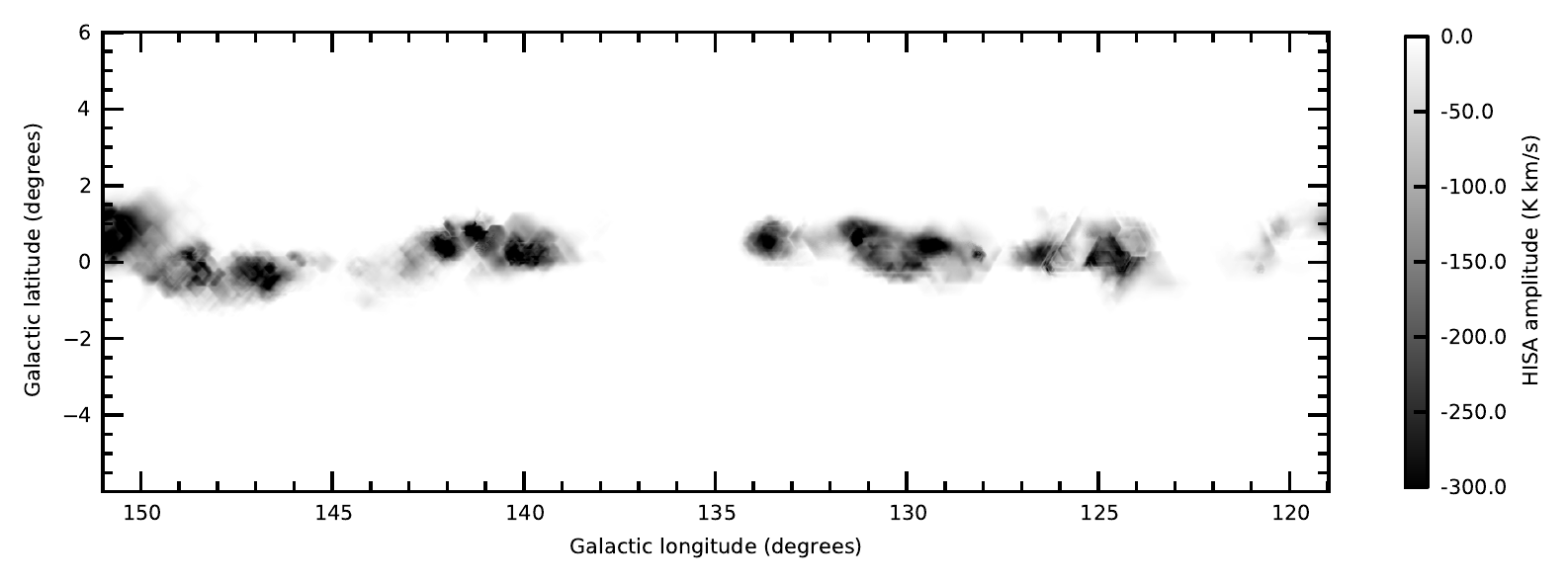}
\label{fig:HISA_Feedback5}}
\subfigure[Feedback10]{
  \includegraphics[scale=1.1]{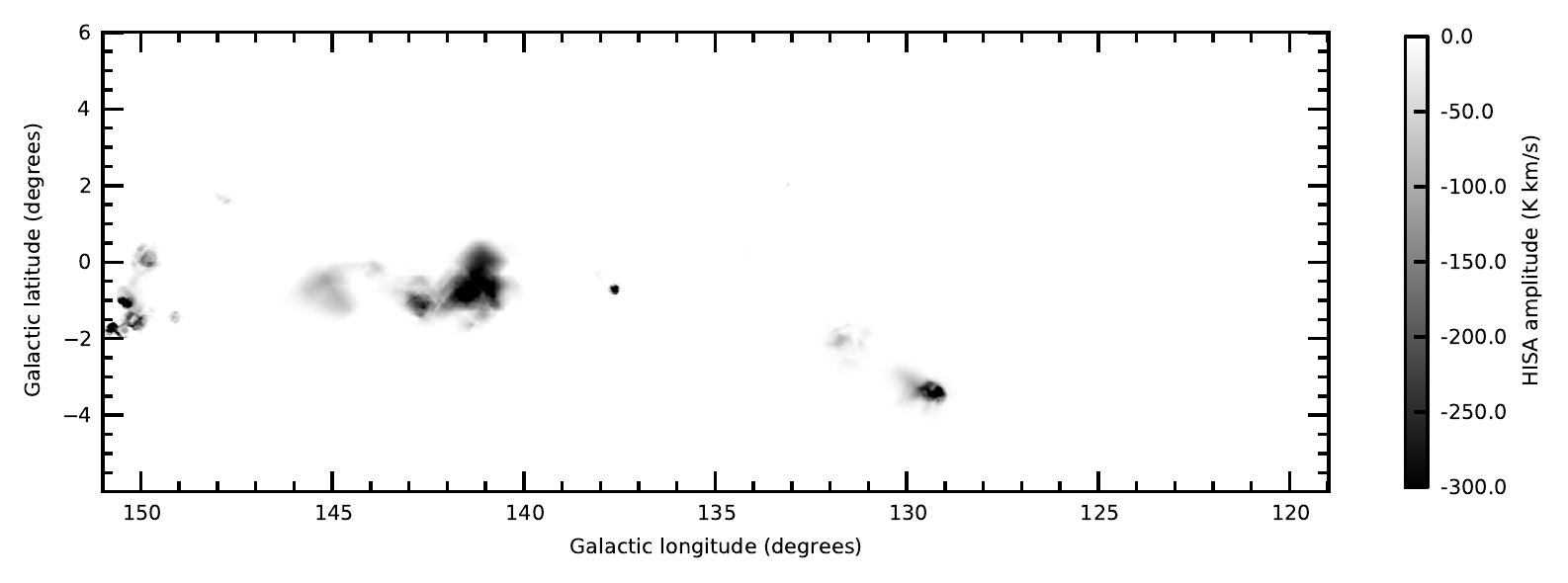}
\label{fig:HISA_Feedback10}}
\caption{H{\sc{i}} self-absorption amplitude, calculated from the
  synthetic observations, where the absorption is integrated over all
  velocity channels. Fig.~\ref{fig:HISA_NoFeedback} is from a
  model galaxy without feedback, Fig.~\ref{fig:HISA_Feedback5} is from
  a model galaxy with feedback and 5~per cent star formation
  efficiency, Fig.~\ref{fig:HISA_Feedback10} is from a model galaxy
  with feedback and 10~per cent star formation efficiency}
  \label{fig:neg_dI}
\end{figure*}

The velocity integrated absorption component is plotted in
Fig.~\ref{fig:neg_dI} for the NoFeedback, Feedback5 and
    Feedback10 models. Equivalent plots from the CGPS data are shown
    in figure~1 of \cite{Gibson05}.

In the NoFeedback model (Fig.~\ref{fig:HISA_NoFeedback}) there is
    a broad band of strong HISA, due to an over concentration of
    H{\sc{i}} in the mid-plane. HISA in the NoFeedback case is 
    confined to $\pm1$~degree from the mid-plane, whereas observed
    HISA has a greater vertical extent. The HISA morphology in the
    Feedback5 model is more realistic, with extended, low intensity
    HISA surrounding knots of stronger absorption. Both the morphology
    and vertical extent of the Feedback5 model are more similar to the
    CGPS observations of \cite{Gibson05}.  Although the Feedback5 HISA
    has less substructure in the strong HISA complexes this is to be
    expected, given the small spatial scale of the observed structures
    relative to the model resolution. In the Feedback10 model the HISA
    has a much reduced diffuse component, compared to the other
    models. \cite{Gibson05} find nearly ubiquitous weak HISA and in
    this regard the Feedback10 model does not match the observations.
    Models with higher star formation efficiencies have less material
    in the cold phase, as shown in fig.~4 of \cite{Dobbs11b}, so
    Feedback10 is expected to have less HISA than Feedback5 as there
    is less cold atomic hydrogen.  We conclude that the Feedback5
    model has a more realistic HISA morphology than either of the
    other models, albeit with limited spatial resolution.

\subsection{Properties of material responsible for HISA}
\label{subsection:HISA_material}

\begin{figure*}
  \centering
\subfigure[Feedback5 number density]{
  \includegraphics[scale=0.31]{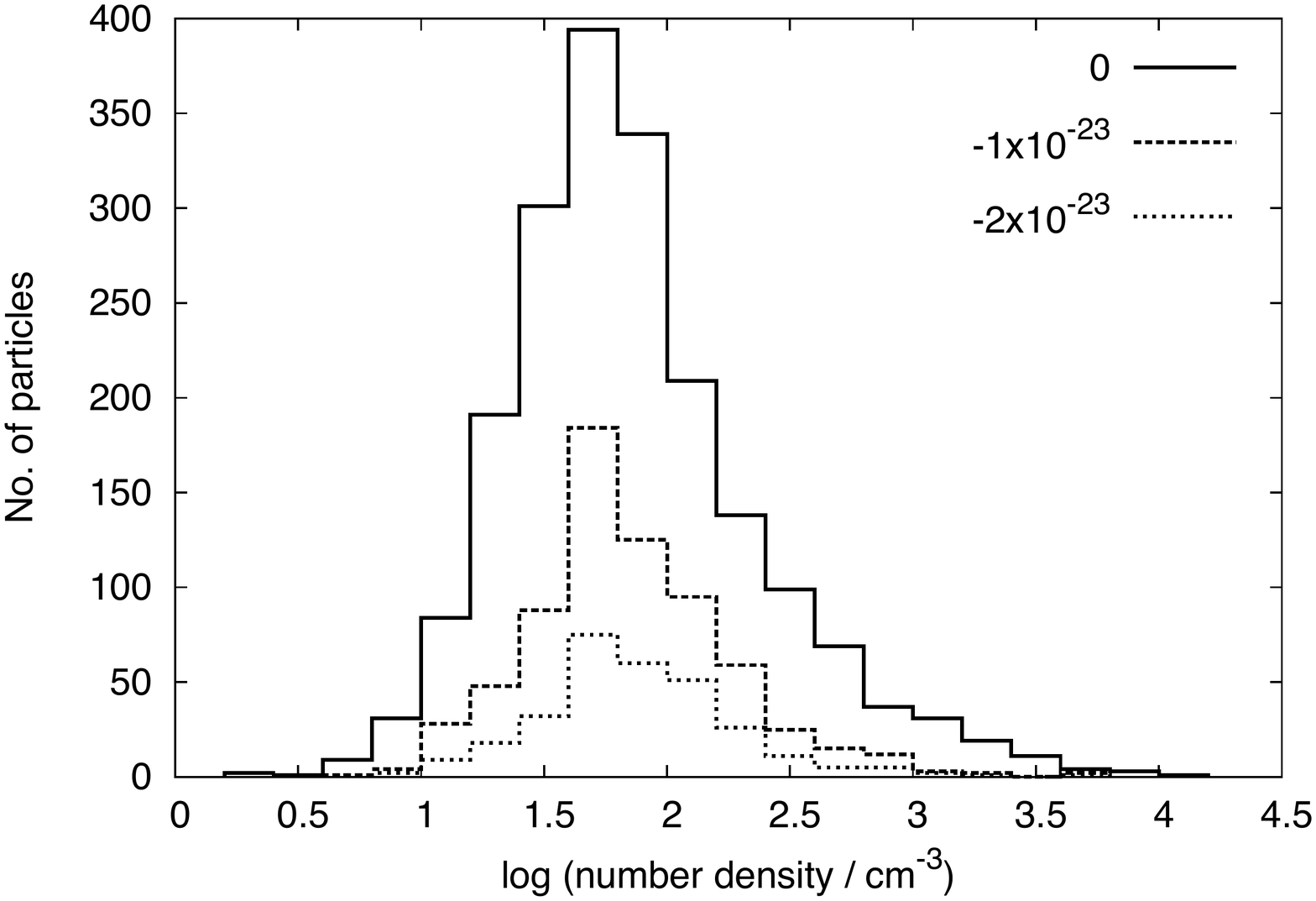}\label{fig:rho_HISA_C}}
\subfigure[Feedback5 temperature]{
  \includegraphics[scale=0.31]{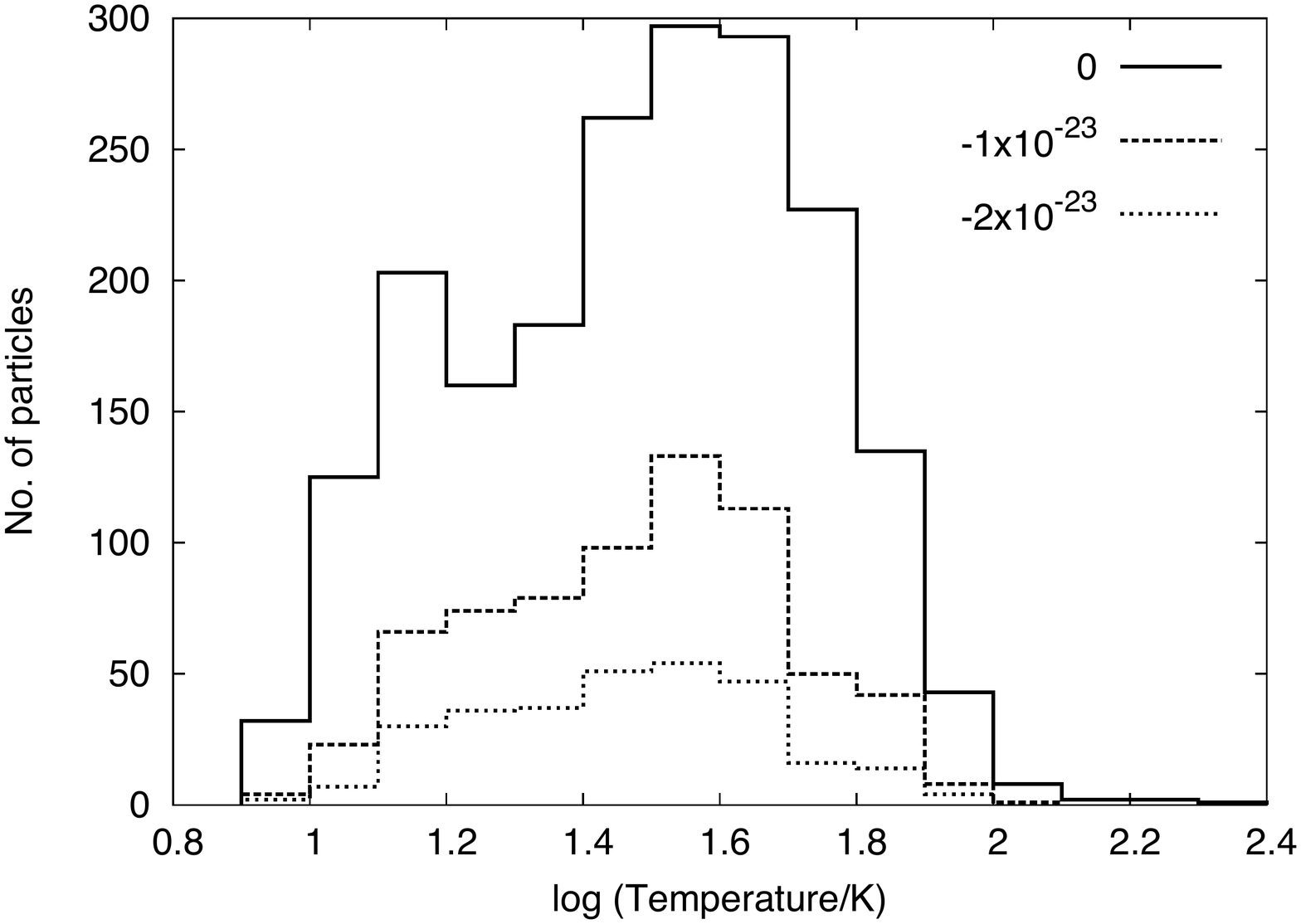}\label{fig:tem_HISA_C}}
\subfigure[Feedback10 number density]{
  \includegraphics[scale=0.31]{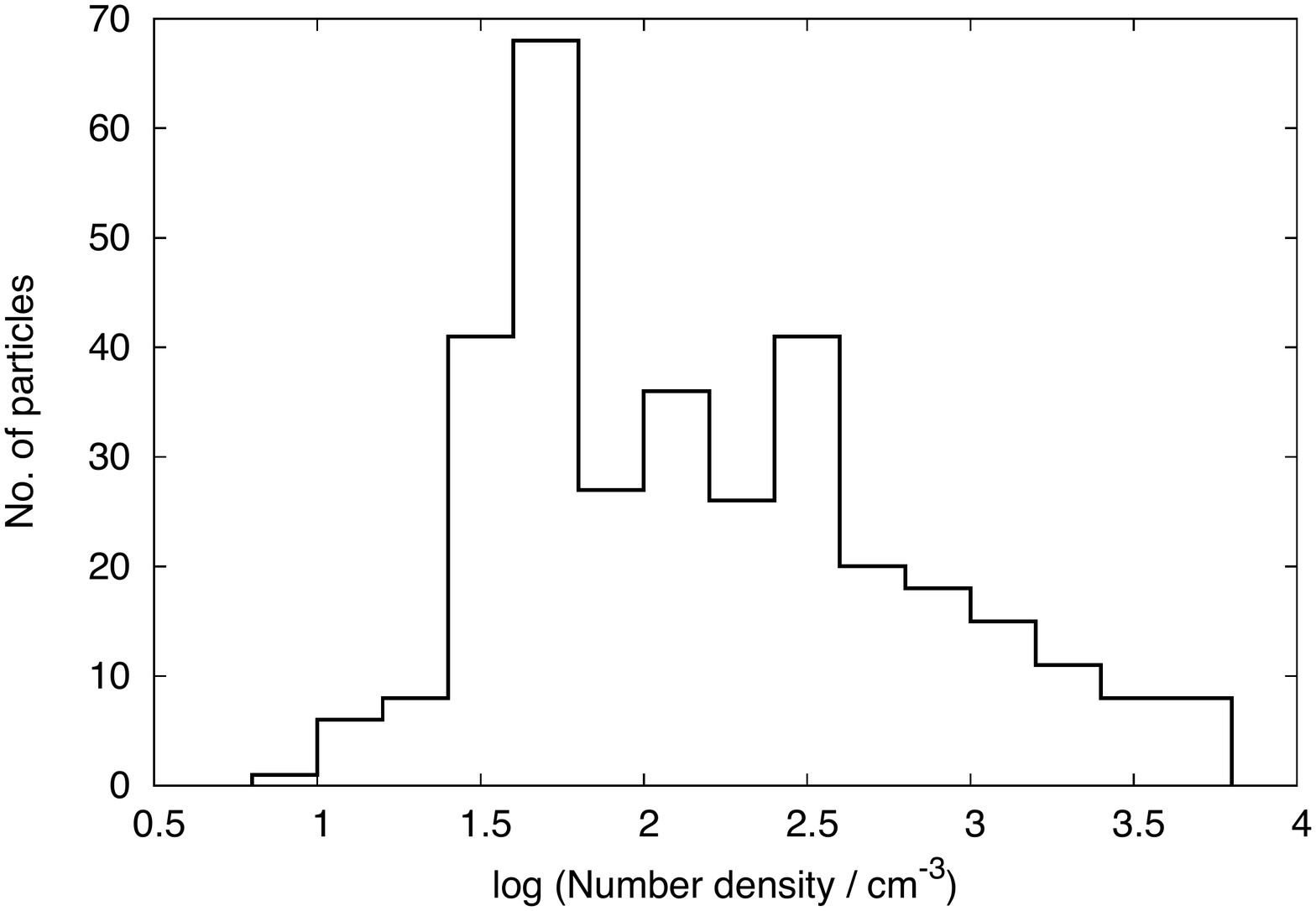}\label{fig:rho_HISA_D}}
\subfigure[Feedback10 temperature]{
  \includegraphics[scale=0.31]{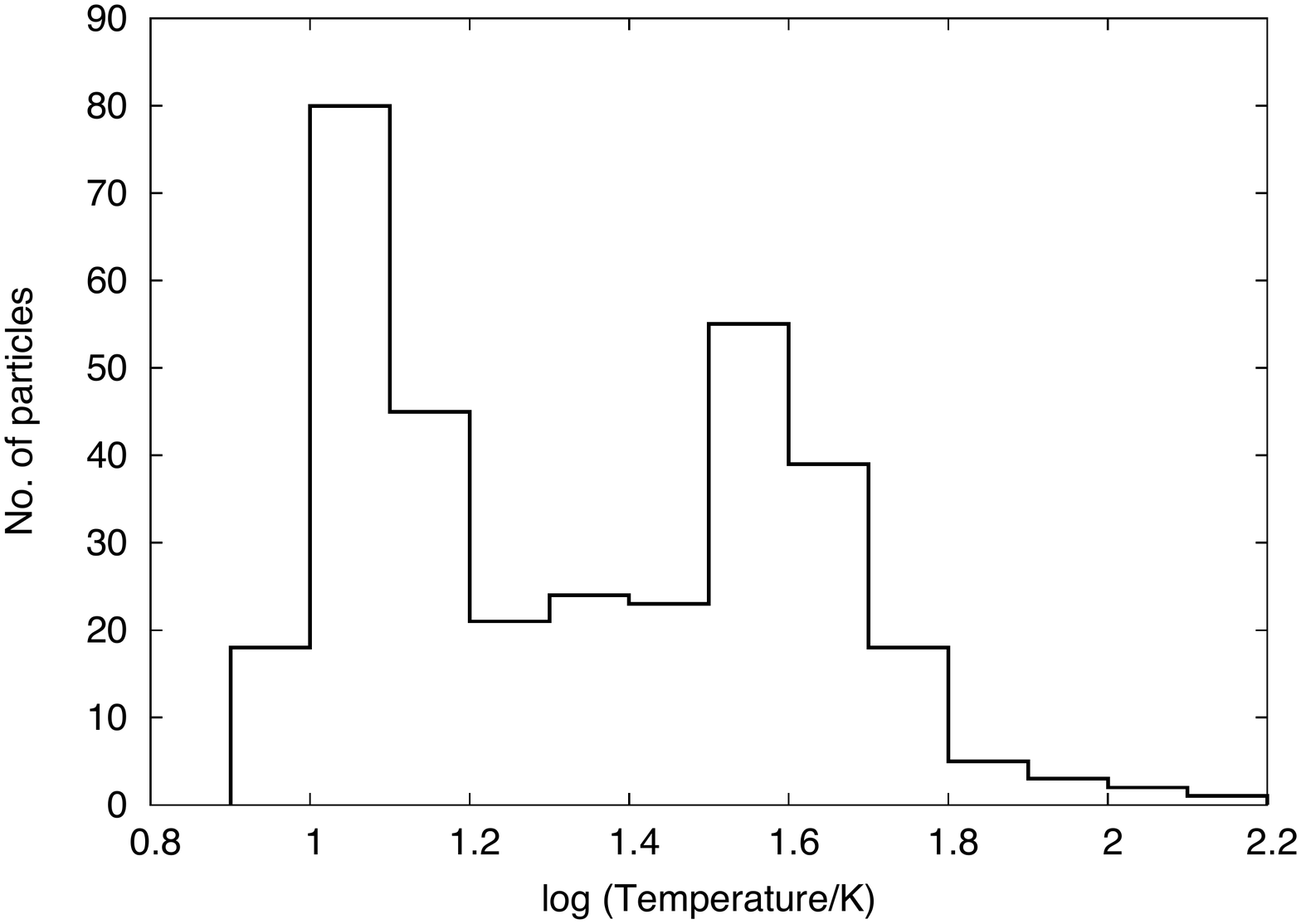}\label{fig:tem_HISA_D}}
  \caption{Number density and temperature distributions of SPH particles
    associated with H{\sc{i}} self-absorption (HISA). For the Feedback5 model three different
    absorption thresholds are used to define HISA particles: $dI<0$
    (solid line), 
    $dI<-1\times10^{-23}$~erg/s/sr (dashed line), and
    $dI<2\times10^{-23}$~erg/s/sr (dotted line). For the Feedback10
    model there are fewer HISA particles and only
    one line is plotted (particles with $dI<0$).} 
  \label{fig:rho_HISA}
\end{figure*}

Figure~\ref{fig:rho_HISA} shows a histogram of number densities
(Fig.~\ref{fig:rho_HISA_C}) and temperatures
(Fig.~\ref{fig:tem_HISA_C}) for absorbing particles from Feedback5, located in the
region used to generate the synthetic survey.  The
solid line is for all particles in a cell with net absorption (6\% of
particles in region), the dashed line is for particles associated with
absorption stronger than $1\times10^{-23}$~erg/s/sr (2\% of particles
in region) and the dotted line is for particles associated with
absorption stronger than $2\times10^{-23}$~erg/s/sr (1\% of particles
in region).
There is no evidence for particles associated with stronger HISA to be
higher density or at a lower temperature, indeed the temperature range
is 10-100K (typical of the cold neutral interstellar medium) for
almost all the absorbing particles. For these histograms the mean
H{\sc{i}} number density is 130--190 $\rm{cm}^{-3}$ and the mean
temperature is 32--36~K.

The density and temperature distributions for absorbing particles from
Feedback10 are shown in Fig.~\ref{fig:rho_HISA_D} and
Fig.~\ref{fig:tem_HISA_D} respectively. Particles associated with net
absorption are plotted as a solid line but no other thresholds are
used, as the number of absorbing particles is much smaller than for
Feedback5 (1\% of particles are absorbing in Feedback10). For
Feedback10 the mean H{\sc{i}} number density is 490~$\rm{cm}^{-3}$ and
the mean temperature is 27K. The range of temperatures and densities
seen in Feedback10 are similar to those seen in Feedback5, although
the HISA in Feedback10 is from slightly colder and denser
material. The star formation efficiency has only a modest effect on
the properties of material seen in HISA, even though the morphology of
the HISA is significantly different.

In their study of HISA clouds in Perseus, \cite{Klaassen05} found
number densities between $100~\rm{cm}^{-3}$ and $1200~\rm{cm}^{-3}$,
and spin temperatures in the range 12K--24K, in an extended HISA
    cloud which they term the ``complex''.  In a smaller,
    isolated HISA feature, which they term the ``globule'', the spin
    temperatures were in the range 8K--22K (no density determination
    was made for the globule.) The globule is a compact structure
    (unresolved in the 1~arcmin main beam of the CGPS) and is much
    smaller than the resolution limit of our synthetic observations
    (see section~\ref{subsection:lb_structure}). The complex region is
    much larger and hence is more like the HISA clouds seen in our
    synthetic data.  The density of HISA producing material in
Feedback5 and Feedback10 is consistent with the density of the complex
region found by \cite{Klaassen05}.  Although the mean temperature of
our HISA producing material is slightly higher than the temperatures
found by \cite{Klaassen05} we note that there is sufficient spread in
our temperature values to encompass their observationally determined
temperatures (see Fig.~\ref{fig:tem_HISA_C} and
Fig.~\ref{fig:tem_HISA_D}). Material can drop out of the gas phase and
form supernovae above a density of $1000~\rm{cm}^{-3}$, so the densest
and coldest regions of HISA producing gas (seen in the observations)
are at the limit of the SPH simulations' representation.

\begin{figure*}
  \centering
\subfigure[HISA particles]{
  \includegraphics[scale=0.31]{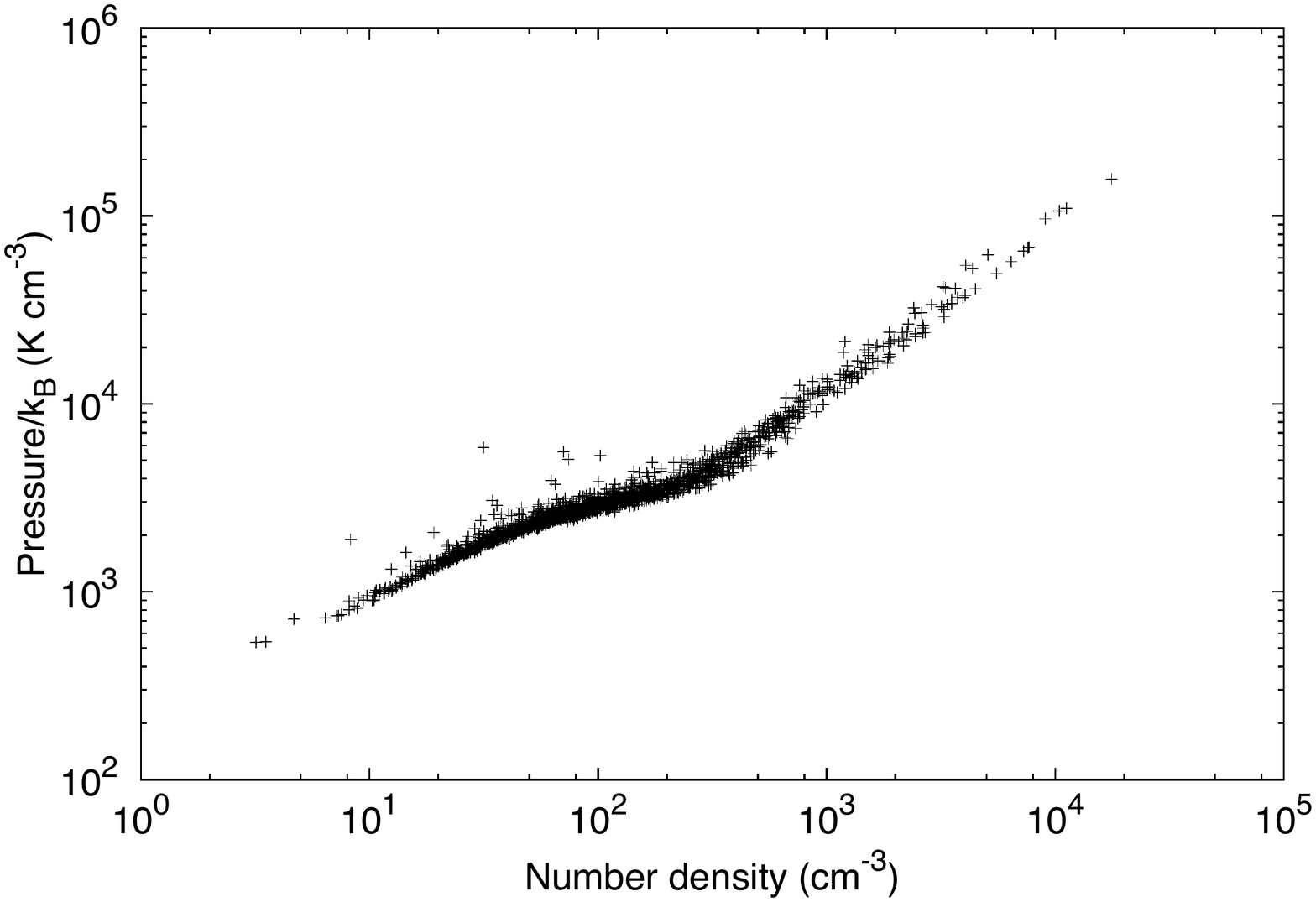}\label{fig:P_vs_n_HISA}}
\subfigure[Other particles]{
  \includegraphics[scale=0.31]{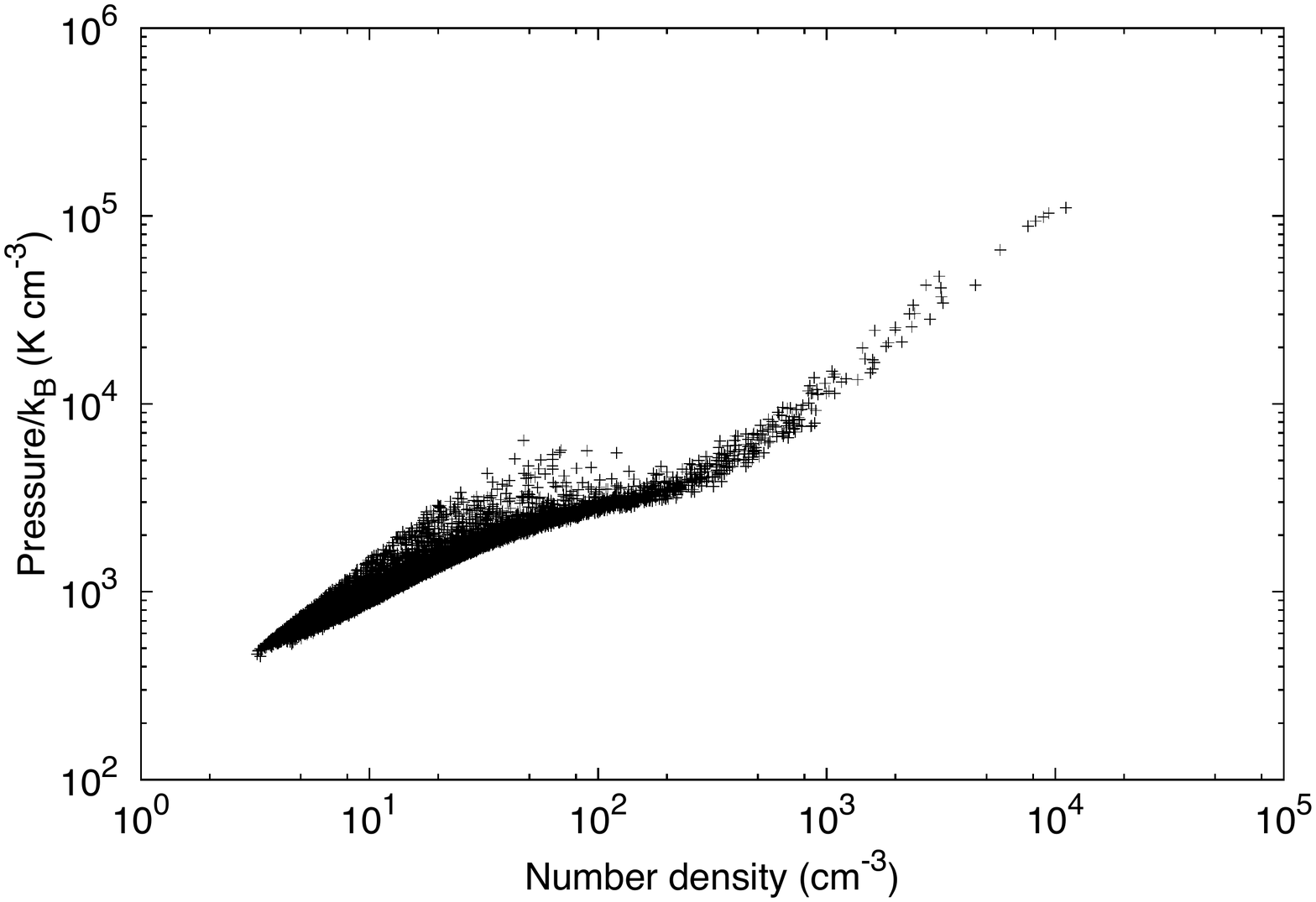}\label{fig:P_vs_n_other}}
  \caption{Plots of pressure against number density for particles
    producing H{\sc{i}} self-absorption (Fig.~\ref{fig:P_vs_n_HISA}) 
  and all other particles with temperatures below 150K
  (Fig.~\ref{fig:P_vs_n_other}), showing that HISA particles are not
  over-pressured compared to other cold particles.} 
  \label{fig:P_vs_n}
\end{figure*}

The next point to be addressed is whether particles which contribute
to HISA are in a different phase to other particles at similar
temperatures, and whether HISA producing material is
in pressure balance with other material. In Fig.~\ref{fig:P_vs_n} we
plot pressure against number density for HISA particles
(Fig.~\ref{fig:P_vs_n_HISA}) and for all other particles with a
temperature below 150K (Fig.~\ref{fig:P_vs_n_other}). There are
approximately five times as many particles in
Fig.~\ref{fig:P_vs_n_other} as in Fig.~\ref{fig:P_vs_n_HISA},
indicating that there is a significant amount of cold, dense H{\sc{i}}
which is not observed in HISA.
For particles with a number density above $10^2 \rm{cm}^{-3}$ there is
no apparent difference between the two populations. At lower densities
the distributions are also similar but there is more spread (towards
higher pressures) in the non-HISA particles. The thermal pressure of
HISA material does not appear to be significantly different to other
material, although we note that the effect of velocity dispersion has
not been included here, which would provide support against gravity in
addition to thermal pressure.

Although the material responsible for HISA is identifiable as the
    cold neutral medium, which is expected to go on to form stars, we
    can be more specific about the fate of the SPH particles
    responsible for HISA.  Particles which are in giant molecular
    clouds (GMCs) are identified, using the clump finding algorithm of
    \cite{DGCK08}, and correlated with whether the material is also
    observed in HISA. In the Feedback5 model 4.0\% of the HISA
    particles are in GMCs, compared to 2.2\% of material in GMCs for
    the galaxy as a whole, and 70\% of HISA material is involved in a
    feedback event (i.e.  involved in star formation) within the next
    20~Myr.  The fraction of HISA material in clouds, and the fraction
    which forms stars, is not found to vary with HISA strength. For
    Feedback10 we find that 2.5\% of HISA material is in GMCs, compared
    to 0.9\% in the galaxy as a whole, and 57\% of HISA material goes on
    to be involved in star formation.

\section{Conclusions and Discussion}
\label{section:conclusions}

The inclusion of feedback in the galaxy models has a significant
    effect on the derived synthetic H{\sc{i}} Galactic plane
    surveys. In the model without feedback (NoFeeback) gas is overly
    confined to the mid-plane, which results in excessive
    absorption. Consequently there are bands of bright emission above
    and below the mid-plane (which are not seen in observations) and
    the vertical extent of the H{\sc{i}} emission is too small. The
    two models which include feedback (Feedback5 and Feedback10) both
    have a larger vertical extent of H{\sc{i}} emission and their
    profiles of H{\sc{i}} emission with latitude match the CGPS
    observations well (provided the normalisation is allowed to
    vary). Based on the vertical distribution of H{\sc{i}} emission
    alone it is not possible to determine whether Feedback5 or
    Feedback10 is more realistic.

  When feedback is included more structure is seen in H{\sc{i}}
  emission, including bubbles of material associated with supernova
  events, however the model does not yet have sufficient spatial
  resolution to capture the very fine scale structure seen in the CGPS
  data. With increases in available computing power it will be
      feasible to run the SPH simulations using more particles, giving
      a higher spatial resolution. This should result in more
      realistic small scale structure in the synthetic
      observations. However if the filamentary structure is due to
  the presence of magnetic fields then these too will need to be
  included before a good match with the observed H{\sc{i}} morphology
  on a small scale can be expected.

  One further respect in which the synthetic data are more realistic
  is the properties of HISA. The NoFeedback model has a broad
      band of strong HISA, unlike HISA seen in observations. When
      feedback is included with 5~per cent star formation efficiency
      the HISA structure is much more like that in the observations of
      \cite{Gibson05}. However if the star formation efficiency is
      increased to 10~per cent there is insufficient HISA, in
      particular there is no ubiquitous weak HISA as found by
      \cite{Gibson05}. Based on the HISA morphology we conclude that a
      5~per cent star formation efficiency is more realistic than a
      10~per cent star formation efficiency.

  With the simulations with feedback the HISA distribution is
  more realistic and it is meaningful to examine the properties of
  material which is seen in HISA. The temperature and density
  properties are confined to a well defined range (typical of cold
  H{\sc{i}}) which is independent of how strongly the material is
  absorbing, and only weakly dependent on the star formation
  efficiency. Although the correspondence between HISA and
      molecular clouds is not one-to-one it is clear that HISA
      preferentially selects material in GMCs. However the nature of
      clouds is dynamic and a given cloud will not be composed of the
      same material as it evolves. Consequently material which is seen
      in HISA at any given time may become part of a GMC later, and
      HISA material currently in a GMC can leave the cloud at a later
      time. Understanding the complex relationships between HISA, GMC
      material and star formation could be facilitated with a time
      series of synthetic observations, which would track particles,
      determine when they are observable as HISA, and correlate this
      with the molecular gas fraction.

\section*{Acknowledgments}

The synthetic survey calculations for this paper were performed on the
DiRAC Facility jointly funded by STFC, the Large Facilities Capital
Fund of BIS, and the University of Exeter. CLD acknowledges funding
from the European Research Council for the FP7 ERC starting grant
project LOCALSTAR. We would like to thank an
anonymous referee for helpful comments.

\bibliographystyle{mn2e}
\bibliography{acreman2011} 

\end{document}